\documentclass[english,aps]{revtex4-1}
\usepackage[T1]{fontenc}
\usepackage[latin9]{inputenc}
\usepackage{amsmath}
\usepackage{graphicx}

\makeatletter

\providecommand{\tabularnewline}{\\}

\usepackage{esdiff}

\makeatother

\usepackage{babel}
\begin{document}

\title{Drift-diffusion current in organic diodes}

\author{Gilles Horowitz}
\email{gilles.horowitz@polytechnique.edu}

\address{CNRS-{}-LPICM, Ecole Polytechnique, 91128 Palaiseau, France}
\begin{abstract}
Because the conductivity of organic semiconductors is very low, a
useful model for the organic diode consists of treating the organic
layer as an insulator, an approximation often referred to as the metal-insulator-metal
(MIM) model. Moreover, the dominant charge carrier injection process
is diffusion, so that a modified Schottky's theory can be used to
derive a simple analytical equation for the current voltage curve
of the diode. Here, we carried out a full analysis of the MIM model
for the organic diode. We show that Schottky's theory is only valid
when charge injection is poor, that is, for high injection barriers.
When the injection barrier is lowered, the current given by Schottky's
theory is still valid in the weak injection regime, when the applied
potential is lower than the diffusion potential. However, it becomes
largely overestimated in the strong injection regime. We also show
that in the strong injection regime, the current given by the MIM
model merges with Mott-Gurnery's space-charge-limited regime.
\end{abstract}
\maketitle

\section{Introduction}

Three mechanisms are usually invoked to rationalize charge carrier
injection in semiconductor diodes: Thermionic emission (TE), the drift-diffusion
(DD) model and tunneling \citep{Sze2006}. TE involves ballistic charge
carrier transport through a depleted (aka space-charge) layer, and
is the model of choice for silicon diodes. Because the mobility in
organic semiconductors is currently several orders of magnitude lower
than that in single crystal silicon, the DD model is generally recognized
as more appropriate to describe the electrical current in organic
diodes.

The DD model rests on two basic equations: Poisson's equation (\ref{eq:poisson})
and the drift-diffusion equation (\ref{eq:DD}).
\begin{align}
\diff[2]Vx & =-\diff Fx=-\frac{qp(x)}{\varepsilon},\label{eq:poisson}\\
j & =qp\mu F-qD\diff px.\label{eq:DD}
\end{align}

Here, the equations are written for positive charge carriers (holes).
$V$ is the electrical potential, $F$ the electrical field, and $p$
the hole density. $\varepsilon$ is the permittivity of the semiconductor,
$j$ the electrical current density, $q$ the elemental charge, $\mu$
the hole mobility and $D$ the hole diffusion coefficient. We will
assume the validity of Einstein's relation, which relates $D$ and
$\mu$ through $D=\mu kT/q$, where $k$ is Boltzmann's constant and
$T$ the absolute temperature.

In spite of they apparent simplicity, the exact resolution of these
equations cannot be fully conducted by analysis; numerical calculations
become a necessity at some stage, which tends to hinder the physical
meaning of the results. Full calculations can be found in papers that
date back to the early days of microelectronics \citep{Shock1953-Space,Skinn1955-Diffusion,Skinn1955-Diffusiona,Wrigh1961-Mechanisms,Bonha1978-Theory},
and in a more recent work by K. Seki \citep{Seki2014-Overall}. At
variance with this analytical approach, the current trend is to perform
numerical resolutions through the finite element method (FEM) \citep{David1997-Device}.
Various commercial packages are available for that purpose. One prominent
advantage of the FEM is that is allows for various refinement in the
calculation, e.g., including unconventional density of states (DOS)
and non constant mobility. However, in spite of they usefulness for
the physical understanding of the process, these simulations are less
appropriate in terms of compact modeling, which requires the development
of simple analytical equations.

The purpose of this paper is to delineates the various options to
analytically resolve the drift-diffusion equation in organic semiconductors,
which are characterized by an extremely low density of thermal charge
carriers.

\section{Theoretical background}

All the equations in this section are written for hole only devices.
The extension to electrons would be straightforward.

\subsection{Schottky's diffusion theory}

The development of this theory can be found in textbooks \citep{Sze2006}.
The principle is to resolve Poisson's and DD equations in sequence.
In the first step, (\ref{eq:poisson}) is used to determine the shape
of the potential in the diode. The result is expressed through the
variation of the valence band edge $E_{v}$ as a function of the distance
$x$ from the metal-semiconductor junction:
\begin{equation}
E_{v}(x)=E_{v}(0)+\frac{q^{2}N_{A}}{\varepsilon}\left(W_{sc}x-\frac{x^{2}}{2}\right),\label{eq:ev_x}
\end{equation}
where :
\begin{equation}
W_{sc}=\sqrt{\frac{2\varepsilon}{qN_{A}}\left(V_{d}-V_{a}-\frac{kT}{q}\right)},\label{eq:wsc}
\end{equation}
is the space charge layer width. $N_{A}$ is the density of dopants
(acceptors for a p-type semiconductor), $V_{d}$ the diffusion (aka
built-in) potential, defined as the difference between the work function
of the metal and that of the semiconductor, and $V_{a}$ the applied
potential.

The current is now established by rewriting (\ref{eq:DD}) as :
\begin{equation}
j=\mu kT\left(\frac{p}{kT}\frac{\mathrm{d}E_{v}}{\mathrm{d}x}-\diff px\right),\label{eq:dd_sch}
\end{equation}
which is next integrated using $\exp(E_{v}/kT)$ as an integrating
factor:
\begin{equation}
j\int_{0}^{W_{sc}}\exp\left(-\frac{E_{v}}{kT}\right)\mathrm{d}x=-\mu kT\left[p\exp\left(-\frac{E_{v}}{kT}\right)\right]_{0}^{W_{sc}}.\label{eq:j_sch}
\end{equation}

Using the Fermi level of the metal as the reference energy, the boundary
conditions are given by:
\begin{align}
E_{v}(0) & =-E_{bp,}\label{eq:ev0}\\
E_{v}(W_{sc}) & =-E_{p}+q(V_{d}-V_{a}),\label{eq:evw}\\
p(0) & =N_{v}\exp\left(-\frac{E_{bp}}{kT}\right),\label{eq:p0}\\
p(W_{sc}) & =N_{v}\exp\left(-\frac{E_{p}}{kT}\right).\label{eq:pw}
\end{align}

\begin{figure}[h]
\centering\includegraphics[width=0.3\columnwidth]{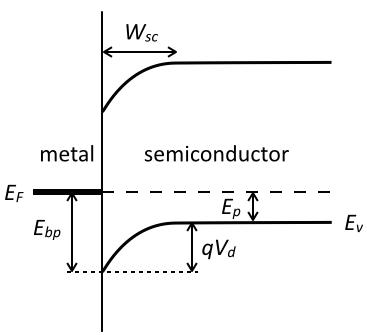}

\caption{\label{fig:schottky_junction}Energy diagram of a Schottky contact
with a p-type semiconductor at equilibrium. Relevant parameters are:
$E_{bp}$: hole energy barrier; $E_{p}$: difference between the Fermi
level and the valence band edge in the bulk of the semiconductor;
$V_{d}$: diffusion (or built-in) potential; $W_{sc}$: space-charge
layer width.}
\end{figure}

$E_{bp}$ is the hole barrier height at the metal-semiconductor interface,
and $E_{p}$ the energy difference between the valence band edge and
the Fermi level in the bulk of the semiconductor. $N_{v}$ is the
effective density of state at valence band edge. An energy diagram
of the junction and the relevant parameters are shown in Fig. \ref{fig:schottky_junction}

Combining all the above leads to:
\begin{align}
j & =j_{d}\left(\exp\frac{qV_{a}}{kT}-1\right),\label{eq:current_sch}\\
j_{d} & =q\mu N_{v}F(0)\exp\left(-\frac{E_{bp}}{kT}\right),\label{eq:jd}
\end{align}
where $F(0)$ is the electric field at the metal-semiconductor interface
($x=0)$.

\begin{figure}[h]
\centering\includegraphics[width=0.6\columnwidth]{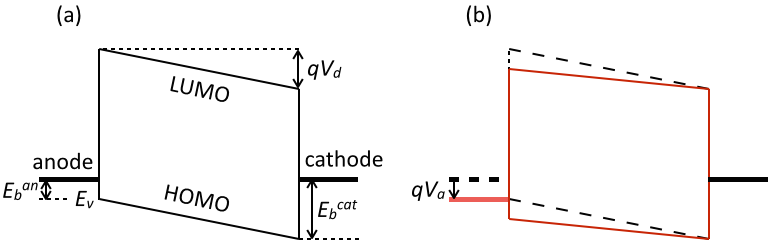}

\caption{\label{fig:mim_junction}Energy diagram of a MIM diode at equilibrium
(a) and low applied voltage (b).}
\end{figure}

In an organic diode, the dopant density and semiconductor thickness
are so small that it is generally accepted that the space charge layer
extends over the whole semiconductor layer, which is referred to as
the full-depletion or metal-insulator-metal (MIM) model. Under such
circumstances, the potential at equilibrium varies linearly with distance,
and the electric field is constant. Attempts to adapt the Schottky
model to such a geometry have been recently made \citep{Nguye2001-influence,Bruyn2013}.
The new energy diagram is shown in Fig. \ref{fig:mim_junction}.

Now we have to consider both sides of the device; the hole injecting
electrode is called anode, and the cathode is hole blocking. Here,
we restrict to a hole only diode, when the electron injection barriers
are so high that only holes can be injected at both electrodes.

The variation of the valence band edge is now given by:
\begin{equation}
E_{v}(x)=E_{v}(0)-q(V_{d}-V_{a})\frac{x}{d},\label{eq:evx_mim}
\end{equation}
where $d$ is the thickness of the semiconductor . $V_{d}$ is the
diffusion potential, that is, the energy difference between the work
function of both electrodes, and $V_{a}$ is the voltage difference
applied between the anode and the cathode. Equation (\ref{eq:j_sch})
must now be integrated over the whole thickness of the semiconductor,
thus leading to:
\begin{equation}
j=q\mu p_{0}\frac{V_{d}-V_{a}}{d}\frac{\exp(qV_{a}/kT)-1}{\exp(qV_{d}/kT)-\exp(qV_{a}/kT)},\label{eq:j_mim}
\end{equation}
where $p_{0}=N_{v}\exp(-E_{b}^{an}/kT)$ is the hole density at the
anode ($x=0$), $E_{b}^{an}$ being the hole injection barrier at
the anode.

The main assumption that leads to (\ref{eq:j_mim}) is that the shape
of the potential profile (quadratic for a Schottky diode, linear for
a MIM diode) remains unchanged when a voltage is applied. This is
basically true in a Schottky diode, as shown in Equation (\ref{eq:ev_x}).
However, as will be shown in the following, the assumption only verifies
in a MIM when the injection barrier at the anode is high.

\subsection{Full MIM model}

We now focus on the following equation, which results of a combination
of (\ref{eq:poisson}), (\ref{eq:DD}) and Einstein's relation:
\begin{equation}
j=\varepsilon\mu\left(F\diff Fx-\frac{kT}{q}\diff[2]Fx\right).\label{eq:full_dd}
\end{equation}

Following earlier works \citep{Wrigh1961-Mechanisms,Bonha1978-Theory},
we will use the dimensionless reduced variables defined as follows:
\begin{equation}
x_{r}=\frac{x}{d},\;V_{r}=\frac{V}{V_{T}},\;F_{r}=\frac{d}{V_{T}}F,\;p_{r}=\frac{qd^{2}}{\varepsilon V_{T}}p,\;j_{r}=\frac{d^{3}}{\varepsilon\mu V_{T}^{2}}j,\label{eq:var_red}
\end{equation}
where $V_{T}=kT/q$. (\ref{eq:full_dd}) now writes:
\begin{equation}
\frac{\mathrm{d}^{2}F_{r}}{\mathrm{d}x_{r}^{2}}-F_{r}\frac{\mathrm{d}F_{r}}{\mathrm{d}x_{r}}+j_{r}=0.\label{eq:full_dd_red}
\end{equation}

\subsubsection{The diode without current}

Integrating (\ref{eq:full_dd_red}) with $j_{r}=0$ leads to:
\begin{equation}
\frac{\mathrm{d}F_{r}}{\mathrm{d}x_{r}}-\frac{F_{r}^{2}}{2}+2g^{2}=0,\label{eq:full_dd_red_zeroj_int}
\end{equation}
where $g$ is an integration constant. The solution of this equation
is given by:

\begin{equation}
F_{r}=-2g\coth\left(gx_{r}+\arg\sinh gx_{r0}\right).\label{eq:Fr_zeroj}
\end{equation}

The electrical potential at a point $x_{r}$ between the anode and
the cathode is obtained by integrating (\ref{eq:Fr_zeroj}) between
the anode ($x_{r}=0$) and $x_{r}$, thus leading to:

\begin{equation}
V_{r}=2\ln\left(\cosh gx_{r}+\sqrt{1+\frac{p_{r0}}{2g^{2}}}\sinh gx_{r}\right).\label{eq:vr_jzero}
\end{equation}

Here, $p_{r0}=p_{r}(0)$ is the reduced density of holes at the anode.
The integration constant $g$ can be calculated by writing that the
reduced potential at the cathode ($x=d,x_{r}=1)$ is equal to to the
reduced diffusion potential $V_{rd}$.

\subsubsection{Solution of the equation with current}

The full DD Equation (\ref{eq:full_dd_red}) has the following analytical
solution:
\begin{equation}
F_{r}(x_{r})=-2\alpha\frac{c_{2}\mathrm{A_{i}^{\prime}}(c_{1}+\alpha x_{r})+\mathrm{B_{i}^{\prime}}(c_{1}+\alpha x_{r})}{c_{2}\mathrm{A_{i}}(c_{1}+\alpha x_{r})+\mathrm{B_{i}}(c_{1}+\alpha x_{r})},\label{eq:Fr_anal}
\end{equation}
where $\alpha=\sqrt[3]{j_{r}/2}$. $\mathrm{A_{i}}$ and $\mathrm{B_{i}}$
are Airy's functions, and $\mathrm{A_{i}^{\prime}}$ and $\mathrm{B_{i}^{\prime}}$
their first derivative; $c_{1}$ and $c_{2}$ are integration constants.
Although (\ref{eq:Fr_anal}) looks analytical, it does not allow for
a direct computation of the current-voltage curve of the diode, because
the integration constants $c_{1}$ and $c_{2}$ must be estimated
for each value of the reduced current $j_{r}$.

From the following relationship between Airy's functions

\begin{align}
\mathrm{A_{i}^{\prime\prime}}(z) & =z\mathrm{A_{i}}(z),\nonumber \\
\mathrm{B_{i}^{\prime\prime}}(z) & =z\mathrm{B_{i}}(z),\label{eq:Aiz}
\end{align}
the reduced hole density $p_{r}=\mathrm{d}F_{r}/\mathrm{d}x_{r}$
can be written as:
\begin{equation}
p_{r}=\frac{F_{r}^{2}}{2}-j_{r}x_{r}-2\alpha^{2}c_{1}.\label{eq:pr_anal}
\end{equation}

Equation (\ref{eq:pr_anal}) can now be used to estimate the integration
constants. In a first step, we write the values of the reduced hole
density at the anode and cathode. Assuming a quasi-equilibrium, we
postulate that these values are those at thermodynamic equilibrium
(no overall current). This leads at the anode ($x_{r}=0$):

\begin{equation}
p_{r0}=2\alpha^{2}\left\{ \left[\frac{c_{2}\mathrm{A_{i}^{\prime}(c_{1})+}\mathrm{B_{i}^{\prime}}(c_{1})}{c_{2}\mathrm{A_{i}(c_{1})+B_{i}(c_{1})}}\right]^{2}-c_{1}\right\} ,\label{eq:pr0}
\end{equation}
or:
\begin{align}
c_{2} & =-\frac{\mathrm{B_{i}^{\prime}}(c_{1})-P_{0}\mathrm{B_{i}(c_{1})}}{\mathrm{A_{i}^{\prime}}(c_{1})-P_{0}\mathrm{A_{i}}(c_{1})},\label{eq:c20}\\
P_{0} & =\pm\sqrt{c_{1}+\frac{p_{r0}}{2\alpha^{2}}}.\label{eq:P0}
\end{align}

A similar equation is obtained at the cathode ($x_{r}=1$):
\begin{align}
c_{2} & =-\frac{\mathrm{B_{i}^{\prime}}(c_{1}+\alpha)-P_{1}\mathrm{B_{i}(c_{1}+\alpha)}}{\mathrm{A_{i}^{\prime}}(c_{1}+\alpha)-P_{1}\mathrm{A_{i}}(c_{1}+\alpha)},\label{eq:c2d}\\
P_{1} & =\pm\sqrt{c_{1}+\frac{j_{r}+p_{r1}}{2\alpha^{2}}}=\pm\sqrt{c_{1}+\alpha+\frac{p_{r1}}{2\alpha^{2}}}.\label{eq:P1}
\end{align}

The sign in front of the square root in (\ref{eq:P0}) and (\ref{eq:P1})
depends on the orientation of the electric field at the anode ($x_{r}=0$)
and cathode ($x_{r}=1$). The constant $c_{1}$ is now obtained by
eliminating $c_{2}$ between (\ref{eq:c20}) and (\ref{eq:c2d}):
\begin{multline}
\left[\mathrm{B_{i}^{\prime}}(c_{1})-P_{0}\mathrm{B_{i}(c_{1})}\right]\left[\mathrm{A_{i}^{\prime}}(c_{1}+\alpha)-P_{1}\mathrm{A_{i}}(c_{1}+\alpha)\right]-\\
-\left[\mathrm{A_{i}^{\prime}}(c_{1})-P_{0}\mathrm{A_{i}}(c_{1})\right]\left[\mathrm{B_{i}^{\prime}}(c_{1}+\alpha)-P_{1}\mathrm{B_{i}(c_{1}+\alpha)}\right]=0.\label{eq:c1}
\end{multline}

The electrical potential is calculated by integrating (\ref{eq:Fr_anal})
between 0 and $x_{r}$:
\begin{equation}
V_{r}=2\ln\frac{c_{2}\mathrm{A_{i}}(c_{1}+\alpha x_{r})+\mathrm{B_{i}}(c_{1}+\alpha x_{r})}{c_{2}\mathrm{A_{i}}(c_{1})+\mathrm{B_{i}}(c_{1})}.\label{eq:vr_anal}
\end{equation}

Replacing $c_{2}$ by its value in (\ref{eq:c20}) leads to:
\begin{multline}
V_{r}=2\ln\pi\{\left[\mathrm{B_{i}^{\prime}}(c_{1})-P_{0}\mathrm{B_{i}(c_{1})}\right]\mathrm{A_{i}}(c_{1}+\alpha x_{r})-\\
-\left[\mathrm{A_{i}^{\prime}}(c_{1})-P_{0}\mathrm{A_{i}(c_{1})}\right]\mathrm{B_{i}}(c_{1}+\alpha x_{r})\},\label{eq:vr_anal_c1}
\end{multline}
where we used the identity $\mathrm{A_{i}}(z)\mathrm{B_{i}^{\prime}}(z)-\mathrm{A_{i}^{\prime}}(z)\mathrm{B_{i}}(z)=1/\pi$.

The applied voltage $V_{a}$ is connected to the reduced potential
at $x_{r}=1$ through $V(d)=V_{T}V_{r}(1)=V_{a}-V_{d}$.

\subsection{Space-charge limited current}

A useful approximation of the DD model was first introduced by Mott
\citep{mott1940}, which consists of neglecting the diffusion component
of the current. The SCLC regime becomes valid at high applied voltage,
and also requires strong charge carrier injection at the anode.

Neglecting the diffusion term leads to the following equation:
\begin{equation}
j=\varepsilon\mu F\diff Fx,\label{eq:sclc}
\end{equation}
which can be integrated to:
\begin{equation}
F^{2}=\frac{2j}{\varepsilon\mu}x+C.\label{eq:sclc_int}
\end{equation}

The integration constant $C$ can be estimated by establishing the
hole density at the anode to $p(0)=p_{0}=N_{v}\exp(-E_{b}^{an}/kT)$:
\begin{align}
p(x) & =\frac{\varepsilon}{q}\diff Fx=\frac{j}{q\mu}\left(\frac{2j}{\varepsilon\mu}x+C\right)^{-1/2},\nonumber \\
p(0) & =\frac{j}{q\mu\sqrt{C}},\nonumber \\
C & =\left(\frac{j}{q\mu p_{0}}\right)^{2}.\label{eq:C}
\end{align}

The potential a point at a distance $x$ of the anode is obtained
by integrating the electric field from the anode to this point:

\begin{equation}
V(x)=\int_{0}^{x}F(t)\mathrm{d}t=\frac{\varepsilon\mu}{3j}\left[\left(\frac{2j}{\varepsilon\mu}x+C\right)^{3/2}-C^{3/2}\right].\label{eq:V_sclc_gene}
\end{equation}

Mott-Gurney's model requires no limitation to charge carrier injection,
so $C\rightarrow0\;(p_{0}\rightarrow\infty)$ and the potential profile
becomes:
\begin{equation}
V(x)=\sqrt{\frac{8j}{9\varepsilon\mu}}x^{3/2}.\label{eq:V_sclc}
\end{equation}

Writing (\ref{eq:V_sclc}) at $x=d$ leads to the well-known equation:
\begin{equation}
j=\frac{9}{8}\varepsilon\mu\frac{\left(V_{a}-V_{d}\right)^{2}}{d^{3}},\label{eq:j_sclc}
\end{equation}
where the applied voltage is defined as $V_{a}=V(d)+V_{d}$.

In the reverse case, when the hole density at the anode becomes low,
we can develop the first term in the bracket in the right side of
Equation (\ref{eq:V_sclc_gene}) to the first power of $x$:
\begin{equation}
V(x)=\frac{\varepsilon\mu}{3j}C^{3/2}\left(\frac{3}{2}\frac{2jx}{\varepsilon\mu C}\right)=\sqrt{C}x,\label{eq:V_limit_inj}
\end{equation}
so the current now writes:
\begin{equation}
j=qp_{0}\mu\frac{V_{a}-V_{d}}{d}.\label{eq:j_limit_inj}
\end{equation}

A similar result is reported in Ref. \citep{Lopez2012-Modeling}.

Note that both (\ref{eq:j_sclc}) and (\ref{eq:j_limit_inj}) are
only valid in the strong injection regime, when $V_{a}>V_{d}$.

\section{Results}

\subsection{Potential profile}

We conducted a numerical resolution of the full MIM model with the
commercial package Mathcad. The first step consisted of calculating
the electric field profile $F(x)$ for given values of the current,
from which the potential and charge carrier density profiles were
obtained through numerical integration and derivation, respectively.
The parameters used for the calculations are gathered in Table \ref{tab:parameters}.

\begin{table}[h]
\caption{\label{tab:parameters}Parameters used for the calculation of the
voltage profile of a MIM diode. $\varepsilon_{0}$ is the permittivity
of free space.}

\centering%
\begin{tabular}{|l|c|}
\hline 
Temperature & $T=300\:\mathrm{K}$\tabularnewline
\hline 
Permittivity & $\varepsilon=4\times\varepsilon_{0}$\tabularnewline
\hline 
Mobility & $\mu=1\:\mathrm{cm^{2}/Vs}$\tabularnewline
\hline 
Density of states at valence band edge & $N_{v}=10^{20}\:\mathrm{cm^{-3}}$\tabularnewline
\hline 
Semiconductor thickness & $d=100\:\mathrm{nm}$\tabularnewline
\hline 
Hole injection barrier at the anode & $E_{b}^{an}=0.1$ or 0.3 eV \tabularnewline
\hline 
Diffusion potential & $V_{d}=0.6\:\mathrm{V}$\tabularnewline
\hline 
\end{tabular}
\end{table}

\begin{figure}[h]
\centering\includegraphics[width=0.9\columnwidth]{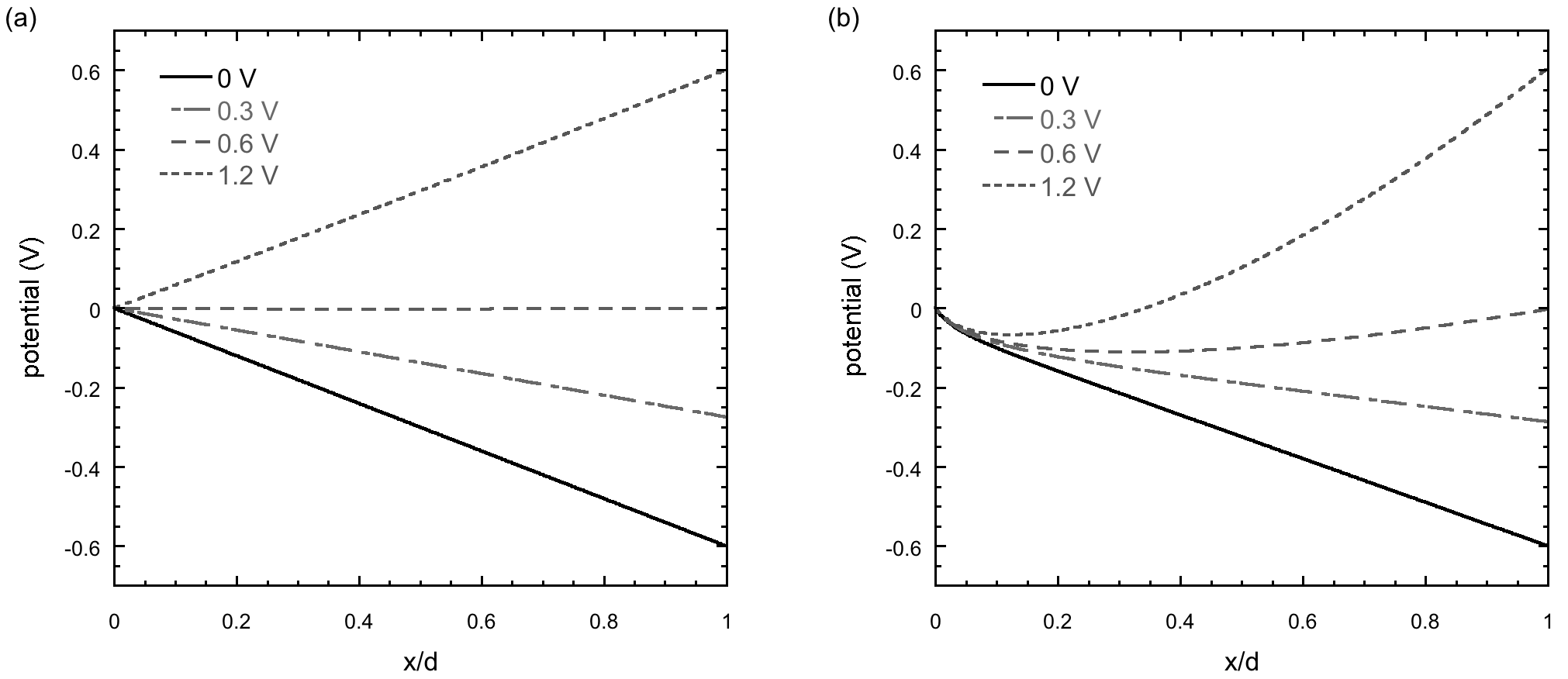}

\caption{\label{fig:V_x}Calculated potential profile of a MIM diode with the
parameters listed in Table \ref{tab:parameters} at various applied
voltages and for an injection barrier at the anode of 0.3 (a) and
0.1 eV (b).}
\end{figure}

The calculated potential profiles of the MIM diode for various values
of the applied voltage are shown in Fig. \ref{fig:V_x}. It clearly
appears that the MIM approximation, in which the potential linearly
varies with distance, is only valid in the case of a high injection
barrier ($E_{b}^{an}=0.3\:\mathrm{eV}$). When the injection barrier
is lower (0.1 eV), a slight curvature appears at the anode ($x=0)$,
which is usually interpreted in terms of accumulation of holes at
this electrode. Moreover, the potential in the direct current regime
($V_{\mathrm{applied}}>V_{d}$) is no longer a straight line; instead,
it presents an upward curvature. Comparing with the voltage profile
described by Equation (\ref{eq:V_sclc}), this can be interpreted
in terms of space charge limited regime, as will be confirmed in the
following.

\subsection{Current-voltage curves}

\begin{figure}[h]
\centering\includegraphics[width=0.5\columnwidth]{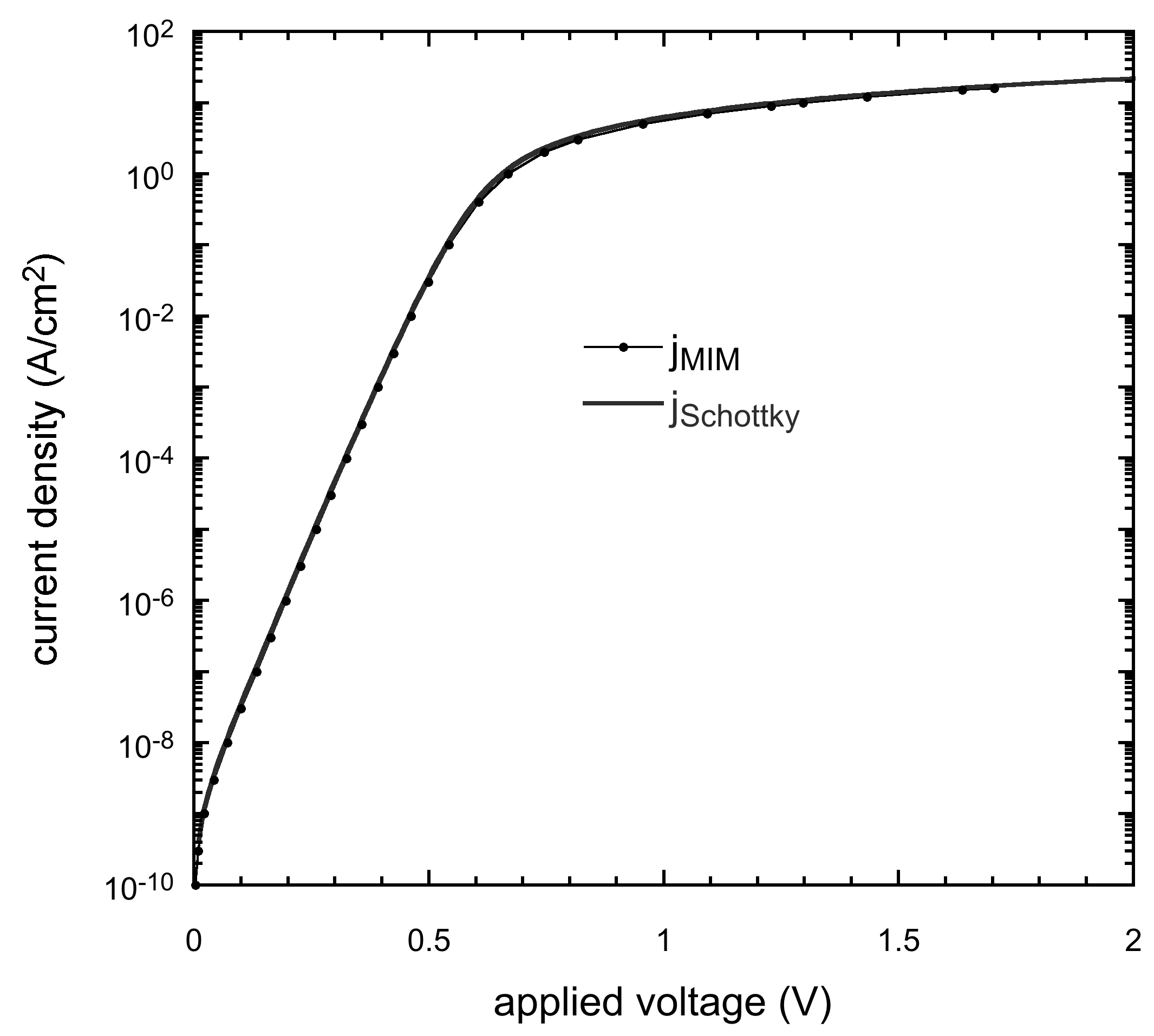}

\caption{\label{fig:iv_van03}IV curve for a MIM diodes with the parameters
in Table \ref{tab:parameters} and a hole injection barrier of 0.3
eV. Filled circles correspond to data numerically calculated from
the full MIM model, and the dashed line to the Schottky theory.}
\end{figure}

\begin{figure}[h]
\centering\includegraphics[width=0.5\columnwidth]{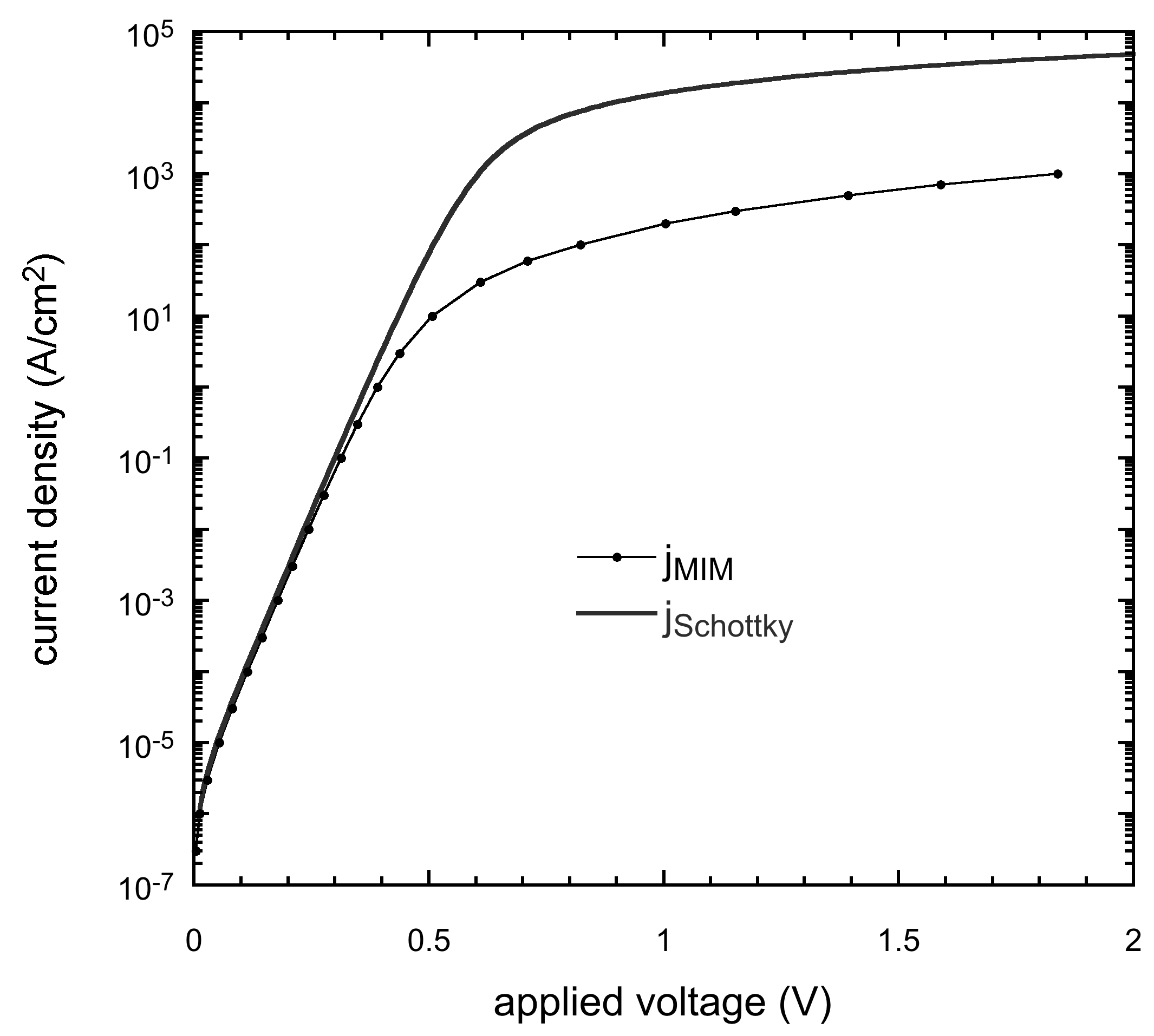}

\caption{\label{fig:iv_van01}Same as Figure \ref{fig:iv_van03} for a hole
injection barrier of 0.1 eV.}
\end{figure}

Calculated current-voltage curves of MIM diodes with an injection
barrier at the anode of 0.3 and 0.1 eV are drawn in semi-log plot
in Figures \ref{fig:iv_van03} and \ref{fig:iv_van01}, respectively.

The exact MIM model is in good agreement with Schottky's theory for
a hole injection barrier of 0.3 eV, that is, when injection efficiency
is poor. When improving charge carrier injection by lowering the injection
barrier down to 0.1 eV, the agreement remains good in the weak injection
regime (applied voltage lower than the diffusion potential $V_{d}$).
However, a discrepancy of nearly two order of magnitude is observed
under strong injection, when the applied voltage is in excess of the
diffusion potential.

\begin{figure}[h]
\centering\includegraphics[width=0.5\columnwidth]{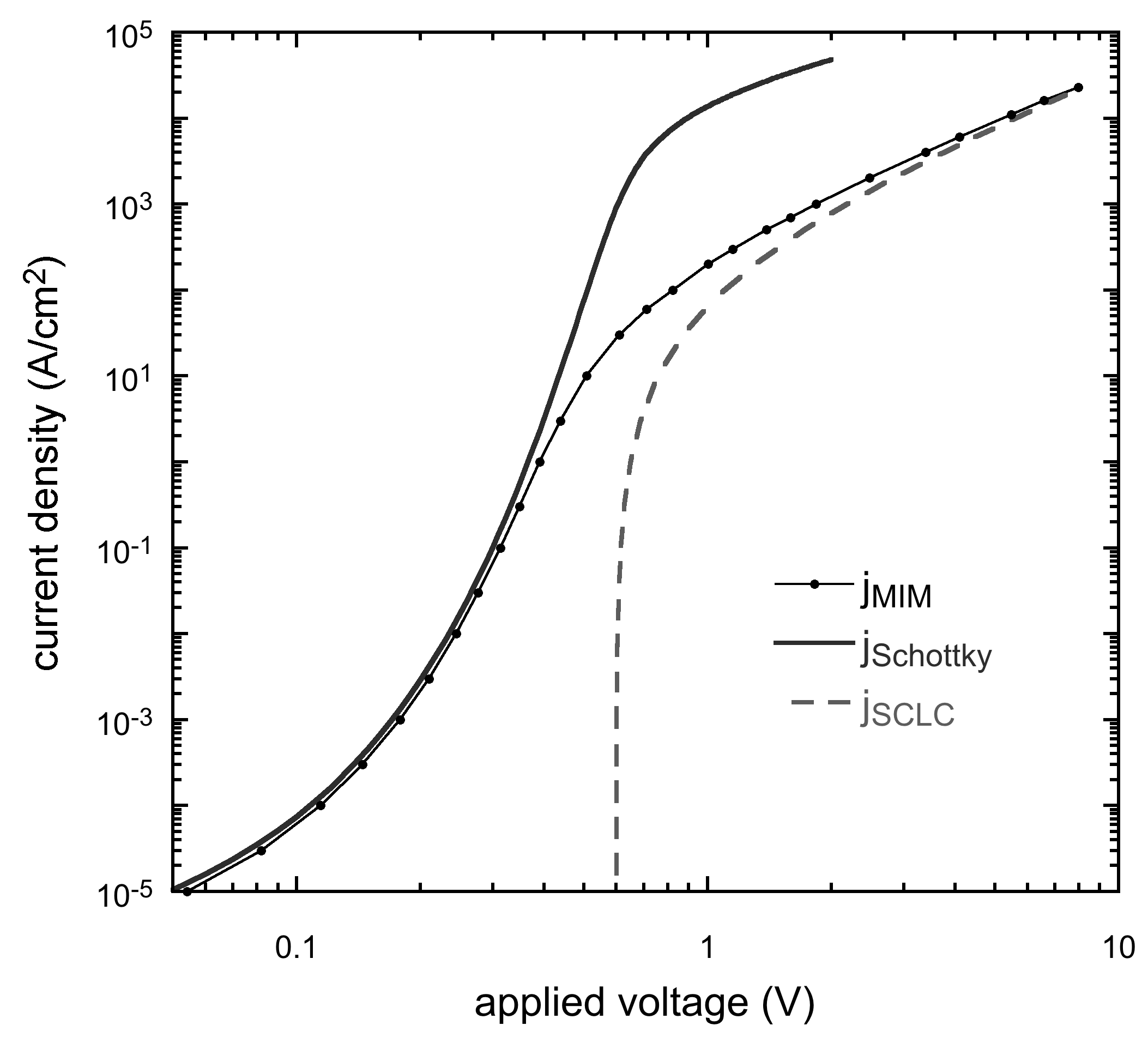}

\caption{\label{fig:iv_loglog_van01}Log-log plot of the current-voltage curve
in Figure \ref{fig:iv_van01} in the strong injection regime ($V_{\mathrm{applied}}>V_{d}$).
The space charge limited current is shown by the dashed line.}
\end{figure}

A log-log plot of the current-voltage curves is shown in Figure \ref{fig:iv_loglog_van01}.
Here, we have also calculated the space charge limited current through
Equations (\ref{eq:C}) and (\ref{eq:j_sclc}). Interestingly, the
exact MIM curve now merges with the SCLC at high voltages (strong
injection regime). We also note that Schottky's current is linear
with the applied voltage in the strong injection regime.

\section{Discussion and Conclusion}

Two distinct cases can be separated, depending on the injection barrier
height: poor injection (high barrier) and good injection (low barrier).
In the former case, the organic semiconductor behaves as a perfect
insulator; the voltage profile remains perfectly linear, including
in the direct bias regime, when current is flowing through the diode.
In this particular case, Schottky's theory leads to a current that
is in perfect agreement with the full MIM calculation in all regimes.
It is worth pointing out that, in agreement with our calculation,
poor injection also prevents SCLC to occur at high applied voltages.
Accordingly, the current at high voltage is proportional to the voltage
rather than the voltage to the square.

For a diode with good hole injection, the agreement of Schottky's
theory with the full MIM model restricts to weak injection, when the
applied voltage is lower than the diffusion potential. At higher voltages,
the current predicted by Schottky's theory in overestimated by a factor
of nearly 100. This discrepancy is accompanied by two important points.
First, apart from a slight curvature near the anode due to hole accumulation,
the voltage profile only remains linear in the weak injection regime
($V_{\mathrm{applied}}<V_{d}$). In the strong injection regime, the
profile presents an upward curvature. This observation can be associated
with the fact that under strong injection, the MIM current merges
with the SCLC regime, with a current that is now proportional to the
voltage to the square.

As a final remark, we note that the model developed here assumes that
the organic semiconductor follows a non-degenerate statistics. This
assumption was used when estimating the density of holes at the electrodes.
We have recently shown that this assumption is not fulfilled in the
case of a Gaussian density of states\citep{Horow2015-Validity}, which
best describes the vast majority of disordered organic solids. Further
work will therefore be necessary to extend the model to a degenerate
statistics.
\begin{acknowledgments}
I am profoundly grateful to Prof. Yvan Bonnassieux, Dr. Chang Hyun
Kim and Sungyeop Jung for their constant support during this work.
\end{acknowledgments}

\appendix*

\section{Approximation at low current}

The use of the exact equations (\ref{eq:c1}) and (\ref{eq:vr_anal_c1})
for the electrical potential and integration constant, respectively,
becomes problematic at low current because the value of the constant
$c_{1}$ becomes positive and large, so Airy's function $\mathrm{B_{i}}$
and its first derivative diverge. To work around this issue, we develop
in this appendix an analytical approximation of the equations at low
current.

First, let us recall the asymptotic form of Airy's functions:

\begin{align}
\mathrm{A_{i}}(z) & \sim\frac{\exp\left(-\frac{2}{3}z^{3/2}\right)}{2\sqrt{\pi}\sqrt[4]{z}},\label{eq:Ai_as}\\
\mathrm{A_{i}^{\prime}}(z) & \sim-\frac{\sqrt[4]{z}}{2\sqrt{\pi}}\exp\left(-\frac{2}{3}z^{3/2}\right),\label{eq:DAi_as}\\
\mathrm{B_{i}}(z) & \sim\frac{\exp\left(\frac{2}{3}z^{3/2}\right)}{\sqrt{\pi}\sqrt[4]{z}},\label{eq:Bi_as}\\
\mathrm{B_{i}^{\prime}}(z) & \sim\frac{\sqrt[4]{z}}{\sqrt{\pi}}\exp\left(\frac{2}{3}z^{3/2}\right).\label{eq:DBi_as}
\end{align}

The asymptotic form of the ratios $\mathrm{A_{i}^{\prime}}(z)/\mathrm{A_{i}}(z)$
et $\mathrm{B_{i}^{\prime}}(z)/\mathrm{B_{i}}(z)$ write:
\begin{align}
\frac{\mathrm{A_{i}^{\prime}}(z)}{\mathrm{A_{i}}(z)} & \sim-\sqrt{z},\label{eq:DAi_Ai}\\
\frac{\mathrm{B_{i}^{\prime}}(z)}{\mathrm{B_{i}}(z)} & \sim\sqrt{z}.\label{eq:DBi_Bi}
\end{align}

We can also write the asymptotic forms of the products of Airy functions
as:
\begin{align}
\mathrm{A_{i}}(z)\mathrm{B_{i}}(z) & \sim\frac{1}{2\pi\sqrt{z}},\label{eq:AiBi}\\
\mathrm{A_{i}^{\prime}}(z)\mathrm{B_{i}}(z) & \sim-\frac{1}{2\pi},\label{eq:DAiBi}\\
\mathrm{A_{i}}(z)\mathrm{B_{i}^{\prime}}(z) & \sim\frac{1}{2\pi},\label{eq:AiDBi}\\
\mathrm{A_{i}^{\prime}}(z)\mathrm{B_{i}^{\prime}}(z) & \sim-\frac{\sqrt{z}}{2\pi}.\label{eq:DAiDBi}
\end{align}

Finally, developing $\left(c_{1}+\alpha x_{r}\right)^{3/2}$ in first
of order of $\alpha/c_{1}$ leads to:
\begin{equation}
\left(c_{1}+\alpha x_{r}\right)^{3/2}\simeq c_{1}^{3/2}+\frac{3}{2}\sqrt{c_{1}}\alpha x_{r},\label{eq:X1_dev}
\end{equation}
so we can write the asymptotic form of Airy function at $c_{1}+\alpha$
as: 
\begin{align}
\mathrm{A_{i}}(c_{1}+\alpha x_{r}) & \sim\mathrm{A_{i}}(c_{1})\exp\left(-\sqrt{c_{1}}\alpha x_{r}\right),\label{eq:AiX1}\\
\mathrm{A_{i}^{\prime}}(c_{1}+\alpha x_{r}) & \sim\mathrm{A_{i}^{\prime}}(c_{1})\exp\left(-\sqrt{c_{1}}\alpha x_{r}\right),\label{eq:DAiX1}\\
\mathrm{B_{i}}(c_{1}+\alpha x_{r}) & \sim\mathrm{B_{i}}(c_{1})\exp\left(\sqrt{c_{1}}\alpha x_{r}\right),\label{eq:BiX1}\\
\mathrm{B_{i}^{\prime}}(c_{1}+\alpha x_{r}) & \sim\mathrm{B_{i}^{\prime}}(c_{1})\exp\left(\sqrt{c_{1}}\alpha x_{r}\right).\label{eq:DBiX1}
\end{align}

\subsection*{Approximate electrical potential}

Using the asymptotic forms of Airy's functions, Equation (\ref{eq:vr_anal_c1})
writes:

\begin{align}
V_{r}(x_{r}) & =2\ln\pi\left\{ \left[\mathrm{B_{i}^{\prime}}(c_{1})-P_{0}\mathrm{B_{i}}(c_{1})\right]\mathrm{A_{i}}(c_{1}+\alpha x_{r})-\left[\mathrm{A_{i}^{\prime}}(c_{1})-P_{0}\mathrm{A_{i}}(c_{1})\right]\mathrm{B_{i}}(c_{1}+\alpha x_{r})\right\} ,\nonumber \\
 & =2\ln\pi\left[\left(\sqrt{c_{1}}-P_{0}\right)\mathrm{B_{i}}(c_{1})\mathrm{A_{i}}(c_{1}+\alpha x_{r})-\left(-\sqrt{c_{1}}-P_{0}\right)\mathrm{A_{i}}(c_{1})\mathrm{B_{i}}(c_{1}+\alpha x_{r})\right],\nonumber \\
 & =2\ln\pi\frac{1}{2\pi}\left[\left(\frac{P_{0}}{\sqrt{c_{1}}}+1\right)e^{\sqrt{c_{1}}\alpha x_{r}}-\left(\frac{P_{0}}{\sqrt{c_{1}}}-1\right)e^{-\sqrt{c_{1}}\alpha x_{r}}\right],\nonumber \\
V_{r}(x_{r}) & =2\ln\left[\cosh\sqrt{c_{1}}\alpha x_{r}+\sqrt{1+\frac{p_{r0}}{2c_{1}\alpha^{2}}}\sinh\sqrt{c_{1}}\alpha x_{r}\right].\label{eq:vr_asymp}
\end{align}

Equation (\ref{eq:vr_asymp}) is similar to the potential at zero
current (\ref{eq:vr_jzero}) where the integration constant $g$ is
replace by $\sqrt{c_{1}}\alpha.$ Hence we deduce that as the current
tends to zero, $c_{1}\alpha^{2}\rightarrow g^{2}$ and $c_{1}$ tends
to infinity.

\subsection*{Approximate equation for the integration constant $c_{1}$ }

At low current, the value of $c_{1}$ becomes large and that of the
current reduces; we can therefore neglect $\alpha$ and rewrite Equation
(\ref{eq:c1}) as follows:

\begin{multline}
\left[\frac{\mathrm{B_{i}^{\prime}}(c_{1})}{\mathrm{B_{i}}(c_{1})}-P_{0}\right]\left[\frac{\mathrm{A_{i}^{\prime}}(c_{1})}{\mathrm{A_{i}}(c_{1})}-P_{1}\right]\mathrm{B_{i}}(c_{1})\mathrm{A_{i}}(c_{1}+\alpha)=\\
=\left[\frac{\mathrm{A_{i}^{\prime}(c_{1})}}{\mathrm{A_{i}}(c_{1})}-P_{0}\right]\left[\frac{\mathrm{B_{i}^{\prime}}(c_{1})}{\mathrm{B_{i}(c_{1})}}-P_{1}\right]\mathrm{A_{i}}(c_{1})\mathrm{B_{i}}(c_{1}+\alpha).\label{eq:c1_anal_1}
\end{multline}

Next, we approximate Equation (\ref{eq:P1}) as:

\begin{equation}
P_{1}\simeq\sqrt{c_{1}}\sqrt{\frac{p_{r1}}{2\alpha^{2}c_{1}}+1}\simeq\sqrt{c_{1}}\left(1+\frac{p_{r1}}{4\alpha^{2}c_{1}}\right).\label{eq:P1_approx}
\end{equation}

At this stage, we also need a development at a higher order of the
asymptotic form of $\mathrm{B_{i}^{\prime}}(z)/\mathrm{B_{i}}(z)$.
A useful form was recently derived Kearney and Martin \citep{Kearn2009-Airy}:
\begin{equation}
\frac{\mathrm{B_{i}^{\prime}}(z)}{\mathrm{B_{i}}(z)}\sim\sqrt{z}-\frac{1}{4z},\label{eq:DBi_Bi_2}
\end{equation}
which leads to the final result:

\begin{equation}
\left(P_{0}-\sqrt{c_{1}}\right)\frac{e^{-\sqrt{c_{1}}\alpha}}{\pi}=\left(P_{0}+\sqrt{c_{1}}\right)\left(\frac{p_{r1}}{\alpha^{2}}+\frac{1}{\sqrt{c_{1}}}\right)\frac{e^{\sqrt{c_{1}}\alpha}}{8\pi c_{1}}.\label{eq:c1_approx}
\end{equation}

\bibliographystyle{apsrev4-1}
\bibliography{biblio_MIM_DD_2015}

\begin{thebibliography}{14}%
\makeatletter
\providecommand \@ifxundefined [1]{%
 \@ifx{#1\undefined}
}%
\providecommand \@ifnum [1]{%
 \ifnum #1\expandafter \@firstoftwo
 \else \expandafter \@secondoftwo
 \fi
}%
\providecommand \@ifx [1]{%
 \ifx #1\expandafter \@firstoftwo
 \else \expandafter \@secondoftwo
 \fi
}%
\providecommand \natexlab [1]{#1}%
\providecommand \enquote  [1]{``#1''}%
\providecommand \bibnamefont  [1]{#1}%
\providecommand \bibfnamefont [1]{#1}%
\providecommand \citenamefont [1]{#1}%
\providecommand \href@noop [0]{\@secondoftwo}%
\providecommand \href [0]{\begingroup \@sanitize@url \@href}%
\providecommand \@href[1]{\@@startlink{#1}\@@href}%
\providecommand \@@href[1]{\endgroup#1\@@endlink}%
\providecommand \@sanitize@url [0]{\catcode `\\12\catcode `\$12\catcode
  `\&12\catcode `\#12\catcode `\^12\catcode `\_12\catcode `\%12\relax}%
\providecommand \@@startlink[1]{}%
\providecommand \@@endlink[0]{}%
\providecommand \url  [0]{\begingroup\@sanitize@url \@url }%
\providecommand \@url [1]{\endgroup\@href {#1}{\urlprefix }}%
\providecommand \urlprefix  [0]{URL }%
\providecommand \Eprint [0]{\href }%
\providecommand \doibase [0]{http://dx.doi.org/}%
\providecommand \selectlanguage [0]{\@gobble}%
\providecommand \bibinfo  [0]{\@secondoftwo}%
\providecommand \bibfield  [0]{\@secondoftwo}%
\providecommand \translation [1]{[#1]}%
\providecommand \BibitemOpen [0]{}%
\providecommand \bibitemStop [0]{}%
\providecommand \bibitemNoStop [0]{.\EOS\space}%
\providecommand \EOS [0]{\spacefactor3000\relax}%
\providecommand \BibitemShut  [1]{\csname bibitem#1\endcsname}%
\let\auto@bib@innerbib\@empty
\bibitem [{\citenamefont {Sze}\ and\ \citenamefont {Kwok}(2007)}]{Sze2006}%
  \BibitemOpen
  \bibfield  {author} {\bibinfo {author} {\bibfnamefont {S.~M.}\ \bibnamefont
  {Sze}}\ and\ \bibinfo {author} {\bibfnamefont {K.~N.}\ \bibnamefont {Kwok}},\
  }\href@noop {} {\emph {\bibinfo {title} {Physics of semiconductor
  devices}}},\ \bibinfo {edition} {3rd}\ ed.\ (\bibinfo  {publisher} {John
  Wiley},\ \bibinfo {address} {Hoboken, New Jersey},\ \bibinfo {year} {2007})\
  p.\ \bibinfo {pages} {832}\BibitemShut {NoStop}%
\bibitem [{\citenamefont {Shockley}\ and\ \citenamefont
  {Prim}(1953)}]{Shock1953-Space}%
  \BibitemOpen
  \bibfield  {author} {\bibinfo {author} {\bibfnamefont {W.}~\bibnamefont
  {Shockley}}\ and\ \bibinfo {author} {\bibfnamefont {R.}~\bibnamefont
  {Prim}},\ }\href {\doibase 10.1103/PhysRev.90.753} {\bibfield  {journal}
  {\bibinfo  {journal} {Physical Review}\ }\textbf {\bibinfo {volume} {90}},\
  \bibinfo {pages} {753} (\bibinfo {year} {1953})}\BibitemShut {NoStop}%
\bibitem [{\citenamefont {Skinner}(1955{\natexlab{a}})}]{Skinn1955-Diffusion}%
  \BibitemOpen
  \bibfield  {author} {\bibinfo {author} {\bibfnamefont {S.~M.}\ \bibnamefont
  {Skinner}},\ }\href@noop {} {\bibfield  {journal} {\bibinfo  {journal} {J.
  Appl. Phys.}\ }\textbf {\bibinfo {volume} {26}},\ \bibinfo {pages} {498}
  (\bibinfo {year} {1955}{\natexlab{a}})}\BibitemShut {NoStop}%
\bibitem [{\citenamefont {Skinner}(1955{\natexlab{b}})}]{Skinn1955-Diffusiona}%
  \BibitemOpen
  \bibfield  {author} {\bibinfo {author} {\bibfnamefont {S.~M.}\ \bibnamefont
  {Skinner}},\ }\href@noop {} {\bibfield  {journal} {\bibinfo  {journal}
  {Journal of Applied Physics}\ }\textbf {\bibinfo {volume} {26}},\ \bibinfo
  {pages} {509} (\bibinfo {year} {1955}{\natexlab{b}})}\BibitemShut {NoStop}%
\bibitem [{\citenamefont {Wright}(1961)}]{Wrigh1961-Mechanisms}%
  \BibitemOpen
  \bibfield  {author} {\bibinfo {author} {\bibfnamefont {G.~T.}\ \bibnamefont
  {Wright}},\ }\href@noop {} {\bibfield  {journal} {\bibinfo  {journal}
  {Solid-State Electronics}\ }\textbf {\bibinfo {volume} {2}},\ \bibinfo
  {pages} {165} (\bibinfo {year} {1961})}\BibitemShut {NoStop}%
\bibitem [{\citenamefont {Bonham}\ and\ \citenamefont
  {Jarvis}(1978)}]{Bonha1978-Theory}%
  \BibitemOpen
  \bibfield  {author} {\bibinfo {author} {\bibfnamefont {J.~S.}\ \bibnamefont
  {Bonham}}\ and\ \bibinfo {author} {\bibfnamefont {D.~H.}\ \bibnamefont
  {Jarvis}},\ }\href@noop {} {\bibfield  {journal} {\bibinfo  {journal} {Aust.
  J. Chem.}\ }\textbf {\bibinfo {volume} {31}},\ \bibinfo {pages} {2103}
  (\bibinfo {year} {1978})}\BibitemShut {NoStop}%
\bibitem [{\citenamefont {Seki}(2014)}]{Seki2014-Overall}%
  \BibitemOpen
  \bibfield  {author} {\bibinfo {author} {\bibfnamefont {K.}~\bibnamefont
  {Seki}},\ }\href {\doibase 10.1063/1.4892987} {\bibfield  {journal} {\bibinfo
   {journal} {Journal of Applied Physics}\ }\textbf {\bibinfo {volume} {116}},\
  \bibinfo {pages} {063716} (\bibinfo {year} {2014})}\BibitemShut {NoStop}%
\bibitem [{\citenamefont {Davids}\ \emph {et~al.}(1997)\citenamefont {Davids},
  \citenamefont {Campbell},\ and\ \citenamefont {Smith}}]{David1997-Device}%
  \BibitemOpen
  \bibfield  {author} {\bibinfo {author} {\bibfnamefont {P.}~\bibnamefont
  {Davids}}, \bibinfo {author} {\bibfnamefont {I.}~\bibnamefont {Campbell}}, \
  and\ \bibinfo {author} {\bibfnamefont {D.}~\bibnamefont {Smith}},\
  }\href@noop {} {\bibfield  {journal} {\bibinfo  {journal} {J. Appl. Phys.}\
  }\textbf {\bibinfo {volume} {82}},\ \bibinfo {pages} {6319} (\bibinfo {year}
  {1997})}\BibitemShut {NoStop}%
\bibitem [{\citenamefont {Nguyen}\ \emph {et~al.}(2001)\citenamefont {Nguyen},
  \citenamefont {Scheinert}, \citenamefont {Berleb}, \citenamefont
  {Br{\"u}tting},\ and\ \citenamefont {Paasch}}]{Nguye2001-influence}%
  \BibitemOpen
  \bibfield  {author} {\bibinfo {author} {\bibfnamefont {P.}~\bibnamefont
  {Nguyen}}, \bibinfo {author} {\bibfnamefont {S.}~\bibnamefont {Scheinert}},
  \bibinfo {author} {\bibfnamefont {S.}~\bibnamefont {Berleb}}, \bibinfo
  {author} {\bibfnamefont {W.}~\bibnamefont {Br{\"u}tting}}, \ and\ \bibinfo
  {author} {\bibfnamefont {G.}~\bibnamefont {Paasch}},\ }\href@noop {}
  {\bibfield  {journal} {\bibinfo  {journal} {Organic Electronics}\ }\textbf
  {\bibinfo {volume} {2}},\ \bibinfo {pages} {105} (\bibinfo {year}
  {2001})}\BibitemShut {NoStop}%
\bibitem [{\citenamefont {de~Bruyn}\ \emph {et~al.}(2013)\citenamefont
  {de~Bruyn}, \citenamefont {van Rest}, \citenamefont {Wetzelaer},
  \citenamefont {de~Leeuw},\ and\ \citenamefont {Blom}}]{Bruyn2013}%
  \BibitemOpen
  \bibfield  {author} {\bibinfo {author} {\bibfnamefont {P.}~\bibnamefont
  {de~Bruyn}}, \bibinfo {author} {\bibfnamefont {A.~H.~P.}\ \bibnamefont {van
  Rest}}, \bibinfo {author} {\bibfnamefont {G.~A.~H.}\ \bibnamefont
  {Wetzelaer}}, \bibinfo {author} {\bibfnamefont {D.~M.}\ \bibnamefont
  {de~Leeuw}}, \ and\ \bibinfo {author} {\bibfnamefont {P.~W.~M.}\ \bibnamefont
  {Blom}},\ }\href {\doibase 10.1103/PhysRevLett.111.186801} {\bibfield
  {journal} {\bibinfo  {journal} {Physical Review Letters}\ }\textbf {\bibinfo
  {volume} {111}},\ \bibinfo {pages} {186801} (\bibinfo {year}
  {2013})}\BibitemShut {NoStop}%
\bibitem [{\citenamefont {Mott}\ and\ \citenamefont {Gurney}(1940)}]{mott1940}%
  \BibitemOpen
  \bibfield  {author} {\bibinfo {author} {\bibfnamefont {N.~F.}\ \bibnamefont
  {Mott}}\ and\ \bibinfo {author} {\bibfnamefont {R.~W.}\ \bibnamefont
  {Gurney}},\ }\href@noop {} {\emph {\bibinfo {title} {Electronic Processes in
  Ionic Crystals}}}\ (\bibinfo  {publisher} {Clarendon Press},\ \bibinfo
  {address} {Oxford},\ \bibinfo {year} {1940})\BibitemShut {NoStop}%
\bibitem [{\citenamefont {L{\'o}pez~Varo}\ \emph {et~al.}(2012)\citenamefont
  {L{\'o}pez~Varo}, \citenamefont {Jim{\'e}nez~Tejada}, \citenamefont
  {L{\'o}pez~Villanueva}, \citenamefont {Carceller},\ and\ \citenamefont
  {Deen}}]{Lopez2012-Modeling}%
  \BibitemOpen
  \bibfield  {author} {\bibinfo {author} {\bibfnamefont {P.}~\bibnamefont
  {L{\'o}pez~Varo}}, \bibinfo {author} {\bibfnamefont {J.~A.}\ \bibnamefont
  {Jim{\'e}nez~Tejada}}, \bibinfo {author} {\bibfnamefont {J.~A.}\ \bibnamefont
  {L{\'o}pez~Villanueva}}, \bibinfo {author} {\bibfnamefont {J.~E.}\
  \bibnamefont {Carceller}}, \ and\ \bibinfo {author} {\bibfnamefont {M.~J.}\
  \bibnamefont {Deen}},\ }\href {\doibase 10.1016/j.orgel.2012.05.025}
  {\bibfield  {journal} {\bibinfo  {journal} {Organic Electronics}\ }\textbf
  {\bibinfo {volume} {13}},\ \bibinfo {pages} {1700} (\bibinfo {year}
  {2012})}\BibitemShut {NoStop}%
\bibitem [{\citenamefont {Horowitz}(2015)}]{Horow2015-Validity}%
  \BibitemOpen
  \bibfield  {author} {\bibinfo {author} {\bibfnamefont {G.}~\bibnamefont
  {Horowitz}},\ }\href {\doibase 10.1063/1.4931061} {\bibfield  {journal}
  {\bibinfo  {journal} {Journal of Applied Physics}\ }\textbf {\bibinfo
  {volume} {118}},\ \bibinfo {pages} {115502} (\bibinfo {year}
  {2015})}\BibitemShut {NoStop}%
\bibitem [{\citenamefont {Kearney}\ and\ \citenamefont
  {Martin}(2009)}]{Kearn2009-Airy}%
  \BibitemOpen
  \bibfield  {author} {\bibinfo {author} {\bibfnamefont {M.~J.}\ \bibnamefont
  {Kearney}}\ and\ \bibinfo {author} {\bibfnamefont {R.~J.}\ \bibnamefont
  {Martin}},\ }\href {\doibase 10.1088/1751-8113/42/42/425201} {\bibfield
  {journal} {\bibinfo  {journal} {Journal of Physics A: Mathematical and
  Theoretical}\ }\textbf {\bibinfo {volume} {42}},\ \bibinfo {pages} {425201}
  (\bibinfo {year} {2009})}\BibitemShut {NoStop}%
\end{thebibliography}%

\end{document}